\documentclass[sigconf]{acmart}

\AtBeginDocument{%
  \providecommand\BibTeX{{%
    Bib\TeX}}}

\copyrightyear{2025}
\acmYear{2025}
\setcopyright{cc}
\setcctype{by-sa}
\acmConference[SC Workshops '25]{Workshops of the International Conference
for High Performance Computing, Networking, Storage and Analysis}{November
16--21, 2025}{St Louis, MO, USA}
\acmBooktitle{Workshops of the International Conference for High Performance
Computing, Networking, Storage and Analysis (SC Workshops '25), November
16--21, 2025, St Louis, MO, USA}
\acmDOI{10.1145/3731599.3767552}
\acmISBN{979-8-4007-1871-7/2025/11}

\usepackage{amsmath,amssymb,amsfonts}
\usepackage{algorithmic}
\usepackage{graphicx}
\usepackage{wrapfig}
\usepackage{textcomp}
\usepackage{hyperref}
\usepackage{xcolor}
\usepackage[acronym]{glossaries}
\makeglossaries
\usepackage[skip=5pt,belowskip=-11pt,font=footnotesize]{caption}
\usepackage{listings}
\definecolor{comment}{RGB}{0,128,0}     
\definecolor{string}{RGB}{255,0,0}      
\definecolor{instruction}{RGB}{0,0,255} 
\definecolor{directive}{RGB}{128,0,128} 
\definecolor{register}{RGB}{128,0,0}    

\def\BibTeX{{\rm B\kern-.05em{\sc i\kern-.025em b}\kern-.08em
    T\kern-.1667em\lower.7ex\hbox{E}\kern-.125emX}}

\newcommand{\myorcid}[1]{%
  \href{https://orcid.org/#1}{\raisebox{-0.2ex}{\includegraphics[height=1.5ex]{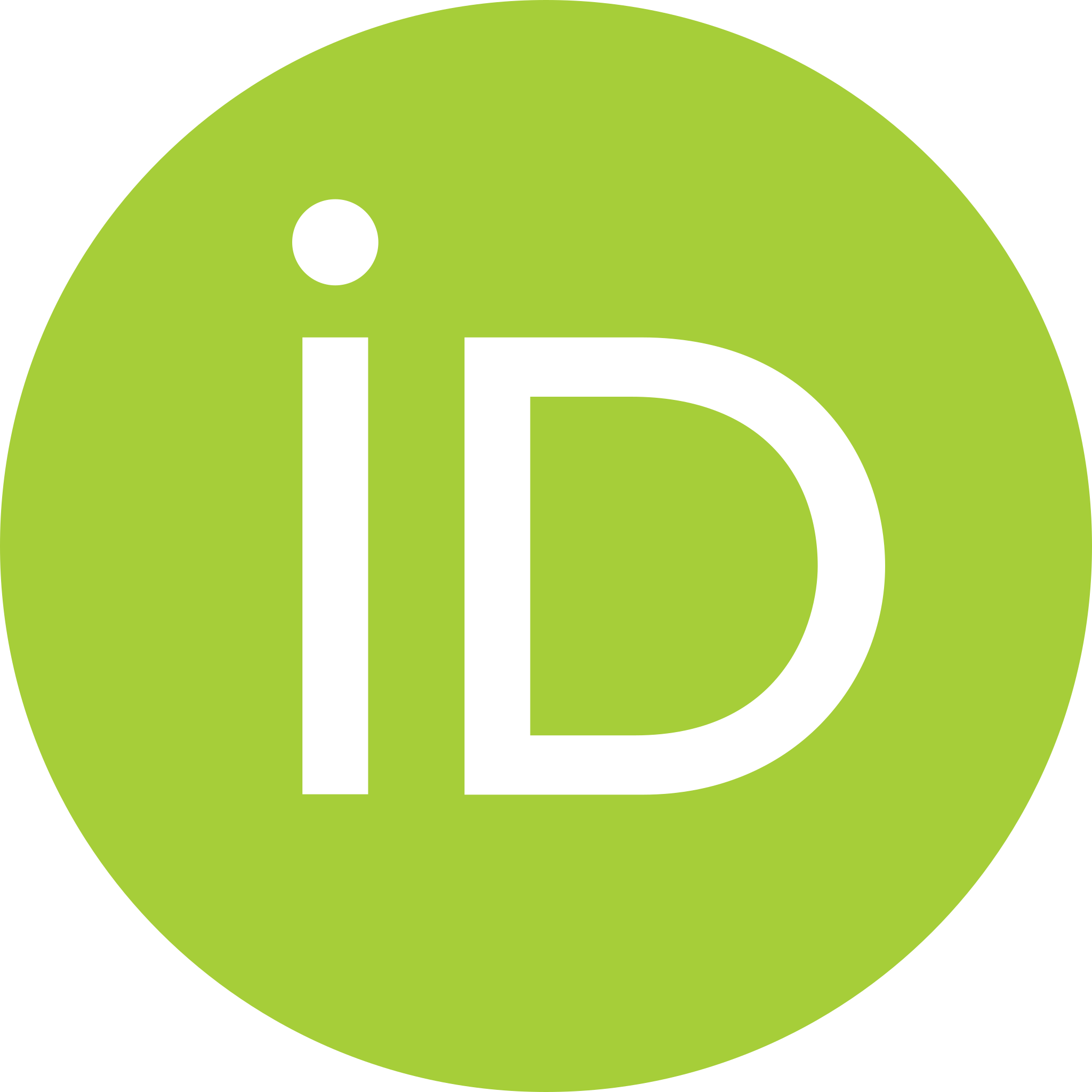}}}%
}

\begin{document}

\title{Tackling the Challenges of Adding Pulse-level Support to a Heterogeneous HPCQC Software Stack}

\subtitle{MQSS Pulse}

\author{Jorge Echavarria \myorcid{0000-0002-3751-5273}}
\affiliation{%
  \institution{Leibniz Supercomputing Centre (LRZ)}
  \city{Garching bei München}
  \country{Germany}}
\email{jorge.echavarria@lrz.de}

\author{Muhammad Nufail Farooqi \myorcid{0000-0002-1609-5847}}
\affiliation{%
  \institution{Leibniz Supercomputing Centre (LRZ)}
  \city{Garching bei München}
  \country{Germany}}
\email{muhammad.farooqi@lrz.de}

\author{Amit Devra \myorcid{0000-0002-7386-1819}}
\affiliation{%
  \institution{Technical University of Munich (TUM)}
  \city{Munich}
  \country{Germany}}
\email{amit.devra@tum.de}

\author{Santana Lujan \myorcid{0009-0005-6237-8302}}
\affiliation{%
  \institution{German Aerospace Center (DLR)}
  \city{Weßling}
  \country{Germany}}
\email{santana.lujan@dlr.de}

\author{Léo Van Damme \myorcid{0000-0002-4311-2497}}
\affiliation{%
  \institution{Technical University of Munich (TUM)}
  \city{Munich}
  \country{Germany}}
\email{leo.van-damme@tum.de}

\author{Hossam Ahmed \myorcid{0009-0000-2498-1198}}
\affiliation{%
  \institution{Leibniz Supercomputing Centre (LRZ)}
  \city{Garching bei München}
  \country{Germany}}
\email{hossam.ahmed@lrz.de}

\author{Martín Letras \myorcid{0000-0002-1429-8982}}
\affiliation{%
  \institution{Leibniz Supercomputing Centre (LRZ)}
  \city{Garching bei München}
  \country{Germany}}
\email{martin.letras@lrz.de}

\author{Ercüment Kaya \myorcid{0000-0001-5073-8159}}
\affiliation{%
  \institution{Leibniz Supercomputing Centre (LRZ)}
  \city{Garching bei München}
  \country{Germany}}
\email{ercuement.kaya@lrz.de}

\author{Adrian Vetter \myorcid{0009-0001-7029-3924}}
\affiliation{%
  \institution{planqc GmbH}
  \city{Garching bei München}
  \country{Germany}}
\email{adrian@planqc.eu}

\author{Max Werninghaus \myorcid{0000-0002-6011-9498}}
\affiliation{%
  \institution{Walther-Meißner-Institut (WMI)}
  \city{Garching bei München}
  \country{Germany}}
\email{max.werninghaus@wmi.badw.de}

\author{Martin Knudsen \myorcid{0009-0000-9752-8829}}
\affiliation{%
  \institution{Walther-Meißner-Institut (WMI)}
  \city{Garching bei München}
  \country{Germany}}
\email{martin.knudsen@wmi.badw.de}

\author{Felix Rohde \myorcid{0009-0008-3391-9052}}
\affiliation{%
  \institution{Alpine Quantum Technologies GmbH
(AQT)}
  \city{Innsbruck}
  \country{Austria}}
\email{felix.rohde@aqt.eu}

\author{Albert Frisch \myorcid{0000-0002-4876-6852}}
\affiliation{%
  \institution{Alpine Quantum Technologies GmbH (AQT)}
  \city{Innsbruck}
  \country{Austria}}
\email{albert.frisch@aqt.eu}

\author{Eric Mansfield \myorcid{0009-0006-1855-6666}}
\affiliation{%
  \institution{IQM Quantum Computers}
  \city{Munich}
  \country{Germany}}
\email{eric.mansfield@meetiqm.com}

\author{Rakhim Davletkaliyev \myorcid{0009-0001-5486-5654}}
\affiliation{%
  \institution{IQM Quantum Computers}
  \city{Helsinki}
  \country{Finland}}
\email{rakhim.davletkaliyev@meetiqm.com}

\author{Vladimir Kukushkin \myorcid{0009-0004-4110-8349}}
\affiliation{%
  \institution{IQM Quantum Computers}
  \city{Helsinki}
  \country{Finland}}
\email{vladimir.kukushkin@meetiqm.com}

\author{Noora Färkkilä \myorcid{0009-0007-0429-8763}}
\affiliation{%
  \institution{IQM Quantum Computers}
  \city{Helsinki}
  \country{Finland}}
\email{noora.farkkila@meetiqm.com}

\author{Janne Mäntylä \myorcid{0009-0008-2246-8313}}
\affiliation{%
  \institution{IQM Quantum Computers}
  \city{Helsinki}
  \country{Finland}}
\email{jmantyla@meetiqm.com}

\author{Nikolas Pomplun \myorcid{0009-0005-2091-6766}}
\affiliation{%
  \institution{German Aerospace Center (DLR)}
  \city{Weßling}
  \country{Germany}}
\email{nikolas.pomplun@dlr.de}

\author{Andreas Spörl \myorcid{0009-0003-0727-440X}}
\affiliation{%
  \institution{German Aerospace Center (DLR)}
  \city{Weßling}
  \country{Germany}}
\email{andreas.spoerl@dlr.de}

\author{Lukas Burgholzer \myorcid{0000-0003-4699-1316}}
\affiliation{%
  \institution{Technical University of Munich (TUM)}
  \city{Munich}
  \country{Germany}}
\affiliation{%
  \institution{Munich Quantum Software Company GmbH (MQSC)}
  \city{Garching}
  \country{Germany}}
\email{lukas.burgholzer@tum.de}

\author{Yannick Stade \myorcid{0000-0001-5785-2528}}
\affiliation{%
  \institution{Technical University of Munich (TUM)}
  \city{Munich}
  \country{Germany}}
\email{yannick.stade@tum.de}

\author{Robert Wille \myorcid{0000-0002-4993-7860}}
\affiliation{%
  \institution{Technical University of Munich (TUM)}
  \city{Munich}
  \country{Germany}}
\affiliation{%
  \institution{Munich Quantum Software Company GmbH (MQSC)}
  \city{Garching}
  \country{Germany}}
\email{robert.wille@tum.de}

\author{Laura B. Schulz \myorcid{0000-0002-4702-3440}}
\affiliation{%
  \institution{Argonne National Laboratory (ANL)}
  \city{Lemont, IL}
  \country{USA}}
\email{schulz@anl.gov}

\author{Martin Schulz \myorcid{0000-0001-9013-435X}}
\affiliation{%
  \institution{Leibniz Supercomputing Centre (LRZ)}
  \city{Garching}
  \country{Germany}}
\affiliation{%
  \institution{Technical University of Munich (TUM)}
  \city{Garching}
  \country{Germany}}
\email{martin.schulz@lrz.de}

\renewcommand{\shortauthors}{Echavarria \textit{et al}.}

\newacronym{api}{API}{Application Programming Interface}
\newacronym{cpu}{CPU}{Central Processing Unit}
\newacronym{eqs3}{EQS3}{European Quantum Systems and Software Summit}
\newacronym{gpu}{GPU}{Graphics Processing Unit}
\newacronym{grape}{GRAPE}{Gradient Ascent Pulse Engineering}
\newacronym{hpc}{HPC}{High Performance Computing}
\newacronym{hpcqc}{HPCQC}{High Performance Computing-Quantum Computing}
\newacronym{ir}{IR}{Intermediate Representation}
\newacronym{isv}{ISV}{Independent Software Vendor}
\newacronym{jit}{JIT}{Just-In-Time}
\newacronym{lrz}{LRZ}{Leibniz Supercomputing Centre}
\newacronym{mlir}{MLIR}{Multi-Level Intermediate Representation}
\newacronym{mqss}{MQSS}{Munich Quantum Software Stack}
\newacronym{mqv}{MQV}{Munich Quantum Valley}
\newacronym{nisq}{NISQ}{Noisy Intermediate-Scale Quantum}
\newacronym{ornl}{ORNL}{Oak Ridge National Laboratory}
\newacronym{qc}{QC}{Quantum Computing}
\newacronym{qdessi}{Q-DESSI}{Quantum Development Environment, System Software \& Integration}
\newacronym{qec}{QEC}{Quantum Error Correction}
\newacronym{qdmi}{\textit{QDMI}}{\textit{Quantum Device Management Interface}}
\newacronym{qir}{QIR}{Quantum Intermediate Representation}
\newacronym{qis}{QIS}{Quantum Instruction Set}
\newacronym{qpi}{\textit{QPI}}{\textit{Quantum Programming Interface}}
\newacronym{qpu}{QPU}{Quantum Processing Unit}
\newacronym{qrm}{\textit{QRM\&CI}}{\textit{Quantum Resource Manager \& Compiler Infrastructure}}
\newacronym{qrmi}{QRMI}{Quantum Resource Management Interface}
\newacronym{qnn}{QNN}{Quantum Neural Network}
\newacronym{tem}{TEM}{Technical Exchange Meeting}
\newacronym{vqe}{VQE}{Variational Quantum Eigensolver}

\begin{abstract}
    We study the problem of adding native pulse-level control to heterogeneous \gls{hpcqc} software stacks, using the \gls{mqss} as a case study. The goal is to expand the capabilities of \gls{hpcqc} environments by offering the ability for low-level access and control, currently typically not foreseen for such hybrid systems. For this, we need to establish new interfaces that integrate such pulse-level control into the lower layers of the software stack, including the need for proper representation.

    Pulse-level quantum programs can be fully described with only three low-level abstractions: \textit{ports} (input/output channels), \textit{frames} (reference signals), and \textit{waveforms} (pulse envelopes).
    We identify four key challenges to represent those pulse abstractions at: the user-interface level, at the compiler level (including the \gls{ir}), and at the backend-interface level (including the appropriate exchange format).
    For each challenge, we propose concrete solutions in the context of \gls{mqss}.
    These include introducing a compiled (C/C++) pulse \gls{api} to overcome Python runtime overhead, extending its LLVM support to include pulse-related instructions, using its C-based backend interface to query relevant hardware constraints, and designing a portable exchange format for pulse sequences.
    Our integrated approach provides an end-to-end path for pulse-aware compilation and runtime execution in \gls{hpcqc} environments.
    This work lays out the architectural blueprint for extending \gls{hpcqc} integration to support pulse-level quantum operations without disrupting state-of-the-art classical workflows.
\end{abstract}



\maketitle

\glsresetall

\section{Introduction}
    The convergence of classical \gls{hpc} and emerging \gls{qc} technology into a unified software stack presents unique opportunities---and challenges---for scientific applications.
    Classical \gls{hpc} infrastructures offer mature tooling for orchestration, data movement, and fault‐tolerance while delivering high performance on classical algorithms.
    Quantum hardware adds to this computational strength with novel computational capabilities for a specific class of quantum algorithms, implemented through coherent control of qubit systems to exploit their special properties, but in turn has to rely on the \gls{hpc} infrastructure for seamless integration and operation.
    This includes the execution of the quantum software stack in general, and quantum compilers in particular.
    However, these have primarily focused on gate‐level abstractions, in particular in the context of \gls{hpcqc}, which may obscure the potentially rich, low‐level dynamics accessible on modern quantum devices, making them inaccessible to hybrid workflows.

    Recent studies similarly emphasize the need to improve low-level quantum control to realize the full potential of quantum computing.
    Smith~\textit{et~al}.~\cite{Smith2022Pulse}, for example, point out that, in addition to better qubit fabrication and algorithms, scaling quantum devices toward fault-tolerance requires refined device-level control.
    This reinforces the consensus that enhancing hardware control mechanisms is critical as quantum systems grow in scale, and that this control must be available in modern \gls{qc} software stacks beyond device-level experimental access:
    a quantum software stack must support it natively and across its entire functionality to enable effective \gls{hpcqc} integration.

    Providing direct pulse-level interfaces, for example, allows high-level classical orchestration to couple seamlessly with the hardware control layer.
    For instance, Delgado and Date~\cite{Delgado2025QuantumReady} describe hybrid \gls{hpcqc} workflows in which classical optimization algorithms compute optimal pulse sequences to steer quantum operations.
    In this way, exposing pulse-level control in the software stack effectively bridges the abstraction gap between classical orchestration and the low-level execution of quantum control pulses.

    In this paper, we identify and address the key challenges of integrating this needed pulse‐level support into a scalable, portable framework that spans both classical and quantum systems.
    We use \gls{mqv}'s \gls{mqss} \cite{Schulz:SRW23,mqss:2025} as a case study to demonstrate our approach.

    \section{Introducing Pulse-level Access}

Pulse-level access enables direct control of the mechanisms used to control the qubits on the respective medium for the chosen quantum technology. This can have different physical manifestations (e.g., via microwaves or lasers), but most cases is the result of translating (higher-level) gate descriptions into some kind of pulse description.

\subsection{Why is Pulse-level Helpful?}\label{subsec:why_pulse}

 Low-level control of these pulses enables a more fine-grained control of the \gls{qc} system, which can have several usage scenarios, including automated calibration and optimal control.
 
    \textbf{Automated Calibration:} Calibration is the systematic, continuous, and iterative process of measuring and compensating for various sources of physical and control errors to ensure that the physical operations performed on qubits match their intended logical definitions.
    Establishing a unified abstraction layer that provides low-level access to diverse quantum hardware platforms for pulse-level control operations enables the management and scheduling of calibration routines. 
    This allows \gls{qc} service providers, like \gls{hpc} centers, to monitor system usage patterns and dynamically schedule calibrations based on anticipated demand.
    This capability enables resource-aware calibration planning, which in turn helps tune the quantum hardware toward fidelity levels that meet operational requirements and align with user workloads as well as any additional priority criteria.
    
    Note that automated calibration routines do not include the calibration of the broader \gls{qc} environment (e.g., cryogenics, vacuum systems, or laser alignment) nor the physical initialization of qubits into a stable, ready state.
    Instead, they focus on fine-tuning system parameters such as control pulse amplitudes, durations, frequencies, and phase alignments to optimize gate fidelity and readout performance.
    These calibration procedures are platform-specific in implementation, but conceptually applicable across multiple \gls{qc} technologies, including superconducting qubits, neutral atoms, and trapped ions.

    \begin{figure}[t!]
        \centering
        \includegraphics[width=0.3\columnwidth]{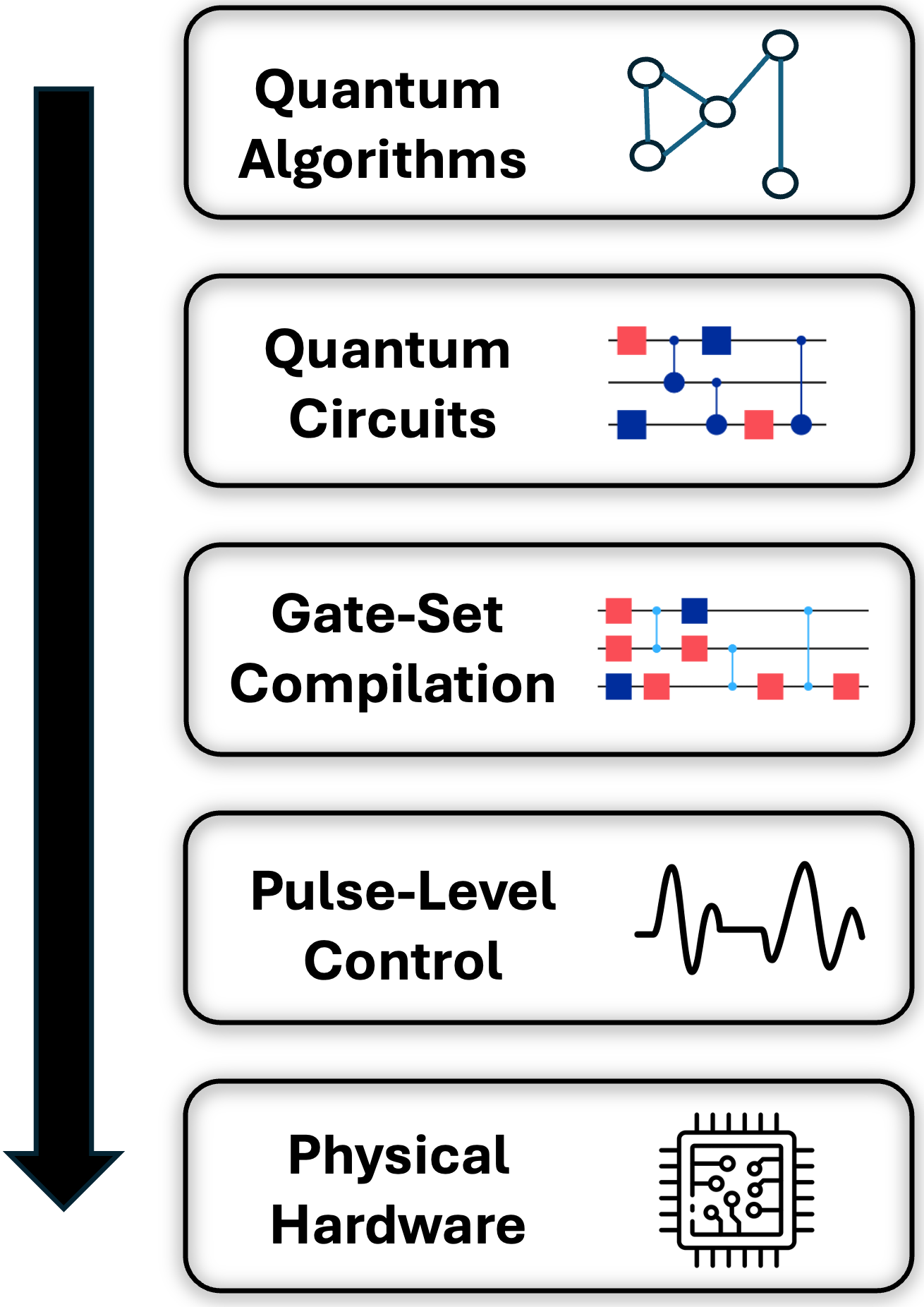}
        \caption{A top-down approach to \gls{qc}: tracing the flow from quantum algorithms and their circuit representations to pulse-level control, i.e., electromagnetic waveforms on target hardware.}
        \label{fig:top_level}
    \end{figure}
    
    For \textit{superconducting qubits}, one of the parameters that demands frequent calibration is the qubit transition frequency, as it can drift on timescales of minutes to hours, therefore requiring continuous real-time tracking via Ramsey-based feedback loops, to ensure the accuracy of the microwave control pulses~\cite{Berritta2025BinarySearch}.
    For \textit{trapped ions systems}, a primary concern is the stability of the electromagnetic trap, with the motional modes frequencies experiencing hour-to-hour drifts of a few hundred hertz and day-to-day drifts of several kilohertz, requiring calibration on these respective schedules~\cite{löschnauer2024scalablehighfidelityallelectroniccontrol}.
    \textit{Neutral atom systems} are dominated by the stability of their laser control systems and the physical integrity of the atom array, which requires calibration of parameters on a minute timescale~\cite{PRXQuantum.6.010101}.
    
    In the \gls{nisq} era, the calibration process is not merely for improving fidelity, but an essential prerequisite for the operation of the quantum accelerator.
    Even \gls{qec} relies heavily on the physical error rates of the underlying physical components to be under a critical threshold of almost 99\% fidelity for single and two-qubit gates to implement a surface code~\cite{old2025faulttolerantstabilizermeasurementssurface}.

    \begin{figure*}[t!]
        \centering
        \includegraphics[width=\textwidth]{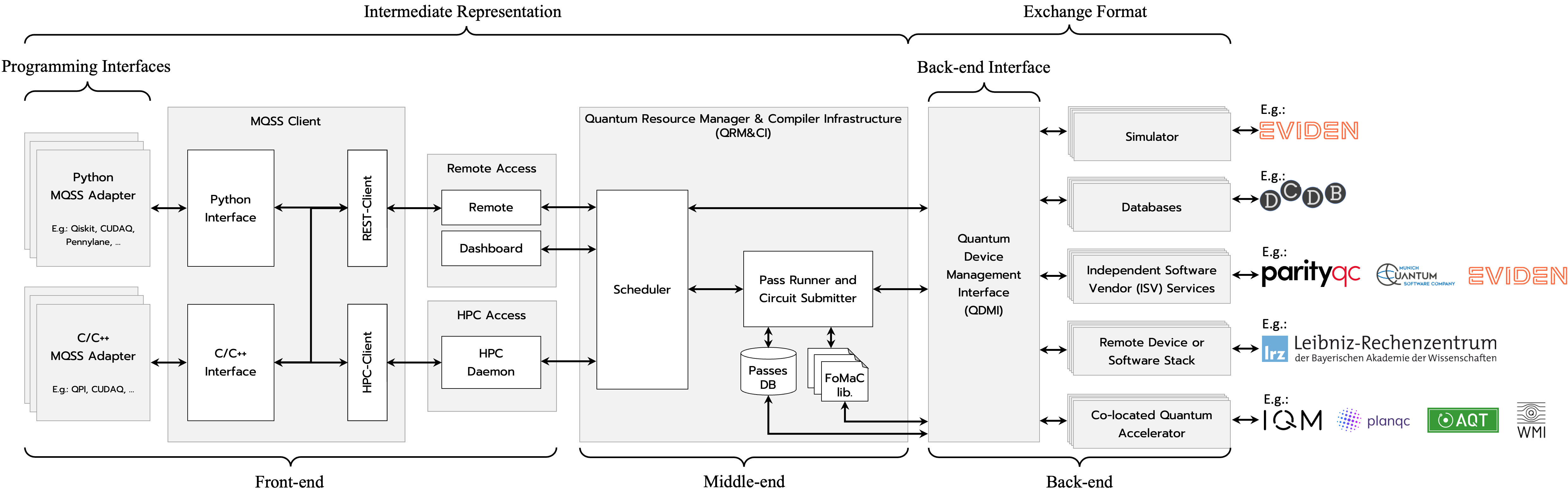}
        \caption{MQSS architecture overview: \textit{MQSS Adapters} (e.g., Qiskit, CUDAQ, PennyLane, and its native C-based QPI) submit gate- and pulse-based jobs to the \textit{MQSS Client}, which handles automatic routing for both local HPC jobs and remote submissions. The \gls{qrm} encompasses MQSS's second-level scheduler, its JIT LLVM-based compiler, and supporting libraries. \textit{QDMI} exposes device capabilities (\textit{ports}, \textit{frames}, \textit{waveforms}, timing/granularity and constraints) to \gls{qrm} during JIT compilation and to the \textit{MQSS Adapters} via the \textit{MQSS Client} during runtime execution. Example \textit{QDMI Devices} shown for illustrating target diversity: classical simulators, databases, ISVs, data centers, and superconducting, neutral atom, and trapped-ion quantum accelerators.}
        \label{fig:mqss_client}
    \end{figure*}
    
    \textbf{Pulse Engineering using Optimal-Control:}
    Designing high-fidelity quantum gates for specific hardware platforms often relies on optimal control techniques to shape control pulses that precisely manipulate qubit dynamics while reducing the impact of errors~\cite{Ansel2024,koch2022quantum,Odelin2019}.
    These pulses are typically engineered to be robust against experimental noise, such as amplitude fluctuations and frequency detuning, which are common in quantum hardware~\cite{Kiely2024,Sorin2024,Nelson2023,Daems2013,Muller2022}.
    In \textit{open-loop} control, pulses are designed offline by simulating the dynamics under a Hamiltonian describing a quantum system, using optimization algorithms such as \gls{grape}~\cite{khaneja2005optimal}. 
    While these methods can be highly effective, they rely on precise knowledge of the system Hamiltonian and may perform poorly if the model does not accurately reflect the true Hamiltonian governing the system dynamics~\cite{Petersen2010,Laflamme2018}.
    In contrast, \textit{closed-loop} control incorporates experimental feedback to iteratively refine pulse parameters based on measured fidelities or system responses, enhancing robustness to hardware imperfections and unmodeled noise~\cite{werninghaus2021leakage,glaser2024sensitivity}.
    A fundamental challenge in closed-loop quantum control is the inability to perform real-time feedback, as quantum measurements irreversibly collapse the system’s state.
    As a result, optimizing control parameters typically requires a large number of repeated experiments, making the process resource-intensive.
    A hybrid approach, which combines open-loop pulse design with closed-loop calibration, is increasingly adopted for achieving near-optimal control on \gls{nisq} devices~\cite{Porotti2023, baum2021experimental,roy2025software}.
    
    \textbf{Pulse-level VQEs:} \glspl{vqe} are widely used hybrid quantum-classical algorithms for estimating ground state energies, but they face serious challenges in the \gls{nisq} era.
    These include decoherence from deep circuits, high classical overhead from frequent recompilation of parameterized Ansätze, and poor trainability due to barren plateaus, where the optimization landscape becomes exponentially flat as system size increases~\cite{mcclean2018barren}.

    An emerging alternative is ctrl-\gls{vqe}~\cite{meiteiGatefreeStatePreparation2021, eggerPulseVariationalQuantum2023}, a pulse-level approach that bypasses traditional gate decomposition and instead optimizes the continuous control waveforms applied to the qubits. 
    This can significantly reduce total circuit duration, thereby mitigating the impact of decoherence, and decrease the energy estimation error compared to gate-based methods.
    Unlike digital \gls{vqe}, ctrl-\gls{vqe} allows the variational optimizer to exploit the full analog nature of the hardware, including access to higher energy states in superconducting qubits (e.g., the $|2\rangle$ level in transmons), which can assist in faster state preparation and enhanced expressivity~\cite{magann2021pulses, sherbertParameterizationOptimizabilityPulselevel2025}.
    These techniques are inspired by and connected to developments in quantum optimal control, where gradient-based pulse shaping is used to achieve high-fidelity target state preparation under physical constraints~\cite{glaser2015training}.

    Although ctrl-\gls{vqe} does not overcome all challenges on the path to quantum advantage, it represents a paradigm shift towards hardware-native quantum algorithm design and may extend to other variational algorithms, such as \glspl{qnn}, especially in scenarios where shorter coherence times and pulse-level customization are critical.

\subsection{Gaining Access to Pulse-level Programming}
Recently, \gls{qc} has seen rapid advancements, with several hardware platforms emerging as viable candidates, each with distinct advantages and limitations. As gate fidelities continue to improve, for approaching and even surpassing error-correction thresholds, pulse-level control is becoming increasingly crucial~\cite{Smith2022Pulse, google2025quantum}. 
It provides fine-grained access to the physical behavior of the qubits, allowing users to optimize performance beyond the limitations of gate-level abstraction. The top-down approach to \gls{qc} shown in Fig.~\ref{fig:top_level} underscores this point by highlighting the critical positioning of pulse-level control in the stack. 

While gate-based programming simplifies algorithm development, it can obscure low-level imperfections such as crosstalk, decoherence, and calibration drift. 
Pulse-level access enables the implementation of a wide range of strategies from the field of quantum optimal control~\cite{Ansel2024,koch2022quantum,Odelin2019}. These include enhancing resilience to experimental imperfections through shaped pulses~\cite{Sorin2024,Kiely2024,Guo2025}, applying dynamical decoupling techniques~\cite{ezzell2023dynamical}, achieving fast and high-fidelity gates~\cite{Jandura2022,Evered2023,VanDamme2025}, and leveraging machine learning–based pulse engineering~\cite{Bukov2018,Porotti2023}. Pulse-level access also enables hardware-aware quantum algorithm development, which is essential for improving fidelity in near-term quantum devices.
As quantum systems scale, pulse-level control will be critical for maximizing the capabilities of increasingly larger \glspl{qpu}~\cite{singkanipa2025demonstration, meiteiGatefreeStatePreparation2021, de2023pulse}.

\section{Case Study: Munich Quantum Software Stack}\label{sec:mqss}
        \Gls{mqss} is an open-source~\cite{mqss-github:25}, \textit{runtime and compilation} full quantum software stack developed as part of the \gls{mqv} initiative~\cite{Schulz:SRW23,mqss:2025,mqss-mqv:25}. 
    We chose \gls{mqss}, as it is openly available on GitHub, accessible to the broader research community, actively maintained via public repositories, as well as already clearly defined hardware/system interface that we can build upon~\cite{mqss-github:25}.
    As shown in Fig.~\ref{fig:mqss_client}, \gls{mqss} provides a modular and extendable infrastructure for hybrid \gls{hpcqc} workflows, offering various programming interfaces, compiler pipelines, and runtime integration layers.
    
    Enabling pulse‑level control in \gls{mqss} involves extending each layer of the stack to recognize and process pulse abstractions.
    At the front end, existing \textit{programming interfaces} must be enhanced to accept pulse descriptions alongside traditional gate-based payloads.
    In the compiler, new \gls{mlir} pass pipelines\footnote{Note that additional pulse‑specific transformations and optimizations can be added incrementally, just as one would extend a conventional LLVM-based gate‑level compiler, such as the \gls{mqss}'s \cite{Swierkowska:ESS24, mqss_mlir:25}.} will lower gate‑based dialects---e.g., Xanadu's Catalyst or NVIDIA's Quake---into a pulse‑oriented dialect---e.g., IBM's \textit{pulse} dialect or a custom equivalent---and subsequently transform that dialect into the proposed pulse \textit{exchange format}.
    Finally, its novel \gls{qdmi}~\cite{Wille:SSESSB24} must be updated to expose and manage the same pulse abstractions used by the \textit{MQSS Adapters} and compiler passes.

\section{Challenges of Adding Pulse-level Support}\label{sec:challenges}
        Built on top of the LLVM framework, \gls{mqss} already provides an extensible multi-dialect compiler infrastructure~\cite{mqss_mlir:25}.
    However, enabling pulse-level support requires further modifications of the following components:

    \begin{itemize}
        \item \textbf{Programming Interfaces}: Both human and automated tools need to have a mechanism to submit pulse-level jobs to the software stack.
        Supporting pulse-level payloads requires updates to both the \textit{MQSS Adapters} and the \textit{MQSS Client}.
        \item \textbf{Intermediate Representation}: A pulse-oriented \gls{mlir} dialect must be developed or adopted to enable the \gls{mqss} compiler to support LLVM compiler passes representing transformations at pulse level.
        \item \textbf{Backend Interface}: \gls{mqss} must be able to query quantum accelerators regarding their supported pulse implementations. 
        The response will inform 1) end-user tools or automated systems via the \textit{MQSS Client}---or any other arbitrary REST-like \gls{api}---2) the compiler for translation and optimization decisions, and 3) runtime or \gls{jit} compilation stages, enabling supportive tooling during compilation and execution.
        \item \textbf{Exchange Format}: The software stack must establish a mutual understanding with quantum accelerators regarding pulse-based payloads. 
        The \gls{qdmi} specification is flexible about the types of payloads it can support, however, we do need to make sure that at least one suitable format exists.
    \end{itemize}

        To achieve consistent pulse support across the stack's front-end, middle-end, and back-end, all components must share a precise, common definition of what constitutes a ``pulse".
    We identify three essential abstractions:

    \begin{itemize}
        \item \textbf{Ports}: A software representation of the hardware input and output channels used to manipulate and read out qubits. It exposes vendor-defined actuation knobs for targeting user-accessible hardware components, such as drive or acquisition channels, while abstracting away device-specific complexity.
        \item \textbf{Waveforms}: A time-ordered array of samples, defining the amplitude envelope of a control signal. The amplitudes can be provided either explicitly or by parametrized functions which, when assigned with specific parameter values, evaluate to a concrete array of samples.
        \item \textbf{Frames}: Stateful timing and carrier signal abstraction combining a reference clock, carrier frequency, and phase. It tracks the elapsed time and provides the timing, frequency, and phase context for playing \textit{waveforms}, enabling precise carrier modulation and virtual phase rotations.
    \end{itemize}

That is, in this paper we treat pulses as control signals with a shape defined by a \textit{waveform}, modulated and timed with a carrier defined by a \textit{frame}, and played on a device component defined by a \textit{port}.

\section{Tackling the Challenges}\label{sec:pulse-level_support}
    By systematically addressing the four challenges identified in Section~\ref{sec:challenges}, a full quantum software stack, such as the \gls{mqss}, can natively orchestrate pulse‑level control across diverse platforms---including ion‑trap, neutral‑atom, and superconducting systems---thereby pre\-par\-ing the stack for future \gls{jit} compilation workflows and hardware-informed pulse‑level optimizations.

    In the following, each challenge is examined in its own sub\-sec\-tion---programming interfaces, \gls{ir}, backend interface, and exchange format---where we detail our strategy for integrating pulse‑level capabilities into the \gls{mqss} and our support of the above abstractions.
    Collectively, these efforts establish a coherent framework for extending the stack toward comprehensive quantum control.

    \subsection{Programming Interfaces}\label{subsec:programming_interfaces}
    Fig.~\ref{fig:mqss_client} shows the \textit{MQSS Client}, which orchestrates quantum jobs via so-called \textit{MQSS Adapters} such as, for example, Qiskit, CUDAQ, PennyLane, and its native \gls{qpi}---a lightweight C-based library designed for \gls{hpcqc} integration~\cite{Kaya:MSFES24}.
Because C/C++ are the dominant languages in \gls{hpc} (powering, for example, CUDA and OpenMP), and because C implementations  far less overhead compared to a scriping language like  Python, we focus on C intefaces as the choice  for large-scale \gls{hpcqc} workloads.
In practice, an \gls{hpc} application invokes \gls{qpi} C functions (embedded in the \textit{MQSS QPI Adapter}) to construct and dispatch quantum programs/kernels. 
The \gls{qpi} library compiles circuits into either an LLVM \gls{ir} (e.g., \gls{qir}) or an \gls{mlir} dialect (e.g., NVIDIA's Quake, Xanadu's Catalyst, or IBM's pulse), and sends them via the \textit{MQSS Client} to target a quantum accelerator.
In this way, \gls{qpi} can be enabled to submit \gls{hpcqc} jobs using any kind of abstraction , including both gate- and potentially pulse-based abstractions, to local or remote quantum hardware through the \gls{mqss} framework.

To add pulse-level control to the enabled abstractions, we extend the \textit{MQSS \gls{qpi} Adapter} with constructs for \textit{ports}, \textit{waveforms}, and \textit{frames},  the three pulse abstractions in our design.
Listing~\ref{lst:qpi-vqe_pulse} shows a simple quantum kernel defined in this extended \gls{qpi} embodying the \textit{pulse-level VQE} use case from Section~\ref{subsec:why_pulse} within function \texttt{pulse\_vqe\_quantum\_kernel}.
Inside an \gls{hpc} loop \gls{qpi} constructs and plays parameterized \textit{waveforms}, collects measurements for energy estimation, and returns control to the classical optimizer for the next iteration.
The snippet begins like a gate-based circuit before defining three \textit{waveforms} (\texttt{qWaveform}) from input amplitudes, playing pulses on specific \textit{ports} (\texttt{qPlayWaveform}), applying \textit{frame} changes (\texttt{qFrameChange}), and finally measuring the results (\texttt{qMeasure}).
The new three \gls{qpi} primitives (i.e., \texttt{qWaveform}, \texttt{qPlayWaveform}, and \texttt{qFrameChange}) operate at native speed due to its C implementation. In detail:

\begin{itemize}
    \item \textbf{\texttt{qWaveform(waveform, amps)}} creates a \textit{waveform} object from given amplitudes (and, implicitly, duration/envelope) to use for subsequent pulses.
    
    \item \textbf{\texttt{qPlayWaveform(port, waveform)}} emits the specified waveform on a hardware \textit{port} (e.g., a qubit drive or coupler channel), physically delivering the pulse to that qubit.
    
    \item \textbf{\texttt{qFrameChange(port, frequency, phase)}} adjusts the carrier \textit{frame} of a qubit \textit{port}, setting its drive frequency and phase offset for precise control.
\end{itemize}

\lstset{
  keywordstyle=\color{blue}\ttfamily,
  commentstyle=\color{comment}\ttfamily,
  stringstyle=\color{teal},
  basicstyle=\ttfamily\footnotesize,
  columns=fullflexible,
  numbers=left,
  numberstyle=\tiny,
  numbersep=4pt,
  frame=lines,
  breaklines=true,
  postbreak=\raisebox{0ex}[0ex][0ex]{\space\ensuremath{\hookrightarrow}},
  showstringspaces=false,
  upquote=true,
  linewidth=\columnwidth,
  xleftmargin=1.1em,
  framexleftmargin=1.1em,
  aboveskip=0.5\baselineskip,
}
\begin{lstlisting}[
    label={lst:qpi-vqe_pulse},
    caption={%
        %Exemplary implementation of a simplified Hardware-Efficient Ansatz~, showcasing some functionality of the proposed pulse-level support in QPI. Note that the \texttt{qPlayDelay} routine is not shown, its proposed function is to schedule a pause on a given \textit{port}.
        %
        Exemplary implementation of a sim\-pli\-fied Hard\-ware-ef\-fi\-cient \textit{Ansatz}~\cite{wang2024HardwareEfficientAnsatz}, showcasing the proposed extension on pulse-level control to the \textit{MQSS QPI Adapter}~\cite{Kaya:MSFES24}. 
        %
        Note that the %\slhat{\texttt{qPlayDelay}}{I would even remove the mention of 'qPlayDelay', as the pulse circuit does not actually require a delay to function and below I have (proposed, see comment) to mention that we only show a subset of the functionalities} %and 
        definition of the \texttt{calculate\_new\_parameters} routine is omitted here, its intended purpose is to %schedule a pause on a given \textit{port} and 
        represent a computationally expensive classical optimization routine. %, respectively. 
        %
        This listing is illustrative and shows only a subset of the proposed functionality. 
        %
        Listings~\ref{lst:ibm_pulse} and \ref{lst:qir_pulse} are intended to be equivalent representations of the same quantum kernel.
    },
    captionpos=b,
    language=c,
    %style=nasm,
    float=tp,
    floatplacement=tbp,
    aboveskip=0.8\baselineskip,
    belowskip=-11pt,
]
#include <mqss_qpi.h>
void pulse_vqe_quantum_kernel(void *results, int nshots, Parameters *p) {
    QCircuit circuit;
    qCircuitBegin(&circuit)
    QClassicalRegisters cr;
    qInitClassicalRegisters(&cr, 2);
    // we begin with X on both qubits
    qX(0);
    qX(1);
    // define the waveforms
    qWaveform(&waveform_1, p->amps_1);
    qWaveform(&waveform_2, p->amps_2);
    qWaveform(&waveform_3, p->amps_3);
    // apply the waves
    qPlayWaveform(qb1_drive_port, waveform_1);
    qPlayWaveform(qb2_drive_port, waveform_2);
    // do the frame changes
    qFrameChange(qb1_drive_port, freq_qb1, p->phase_qb1);
    qFrameChange(qb2_drive_port, freq_qb2, p->phase_qb2);
    // apply the entangling pulse
    qPlayWaveform(qb1_qb2_coupler_port, waveform_3);
    // measure
    qMeasure(0, 0);
    qMeasure(1, 1);
    qCircuitEnd();
    if(!qExecute(dev, circuit, nshots))
        QuantumResult* results = qRead(circuit);
    
    qCircuitFree(circuit);
}
int main(){
    do{
        void* results = malloc(size);
        pulse_vqe_quantum_kernel(&results, nshots, &parameters);
        parameters = calculate_new_parameters(&results, parameters)
    }while( stop_condition == false );
    return 0;
}
\end{lstlisting}

These extensions satisfy our \gls{hpc}-focused design requirements: by implementing them in C, we keep latency low and resource usage minimal, addressing the performance and integration challenges identified earlier. 
Importantly, our approach remains compatible with other pulse-level \glspl{api}.
For example, IBM has deprecated pulse-level support in Qiskit in early 2025, underscoring the need for alternative interfaces.
In contrast, the \textit{MQSS QPI Adapter} is expected to continue to support full analog control in \gls{hpc} environments.
Moreover, the \gls{qpi} abstractions align closely with industry standards: Amazon Braket's Pulse feature, for example, exposes \textit{ports}, \textit{frames}, and \textit{waveforms} through its SDK (including OpenQASM~3.0 support), while the OpenQASM~3.0 specification defines calibration (\texttt{cal}) blocks that explicitly use the same three abstractions.
In practice, this means that a \gls{qpi} pulse program could be translated or interfaced with Braket- or OpenQASM3-style schedules, if desired.

\subsection{Intermediate Representation}\label{subsec:intermediate_representation}
    As described in Section~\ref{sec:mqss}, the \textit{\gls{mqss} Compiler} is fully based on LLVM-\gls{ir}~\cite{Swierkowska:ESS24} and LLVM-\gls{mlir}~\cite{mqss_mlir:25}, where all gate-based quantum circuit transformations are implemented as either \gls{qir} or \gls{mlir} (e.g., NVIDIA's Quake and Xanadu's Catalyst) passes. 
LLVM's built-in pass manager supports \gls{mlir} dialect-agnostic orchestration by allowing both operation-specific and operation‑agnostic passes to be registered and executed on \gls{ir} modules, regardless of the dialect they belong to---as long as the pass is targeted to the correct dialect context.
Thus, any \gls{mlir} job loaded into memory can be processed by a pass suite appropriate for its dialect—simplifying support for new domains like pulse-level control.
This design allows \gls{mqss} to handle multiple dialects---existing or future---without fundamental changes to the pass orchestration system.

For example, IBM's Quantum Engine defines a Pulse \gls{mlir} dialect, where each high-level gate or measurement is lowered into a pulse sequence via provided ``calibration'' \textit{waveforms}~\cite{ibm-qe_compiler:24}.
In this dialect, every gate has an associated pulse \textit{waveform}.
Listing~\ref{lst:ibm_pulse} illustrates an example with this \gls{mlir} dialect.
Three \textit{waveforms} are defined with \path{pulse.def @waveform_*}.
The \path{@pulse_vqe_quantum_kernel} sequence then applies \path{pulse.standard_x} gates (X on each qubit), followed by \path{pulse.play} calls that apply the predefined \textit{waveforms} on drive and coupler \textit{frames}.
It also utilizes \path{pulse.frame_change} to adjust phases/frequencies, and ends with the measurement: a readout \path{pulse.play} on each qubit followed by \path{pulse.capture}, returning the classical bits.
As seen, LLVM-based quantum compilers, such as the one in \gls{mqss}, can natively support gate-level operations and pulse instructions even in one single \gls{ir}.

This \gls{mlir} dialect is compatible with the three key abstractions---\textit{ports}, \textit{frames}, and \textit{waveforms}---as \gls{mlir} constructs aligned with the three core definitions we introduced in Section~\ref{sec:challenges}.
Specifically, \textit{ports} model hardware-specific I/O channels---e.g., actuation or readout interfaces defined by the vendor.
\textit{Waveforms} describe signal envelopes created via \path{create_waveform} operations and emitted through \path{play} operations on mixed frames—structures mixing port channel and frame state.
Further, \textit{frames} combine a logical clock---time that increments with use---with stateful carrier signal parameters---frequency and phase.

\input{listings/circuit_pulse.ll}

Gate-level operations have direct pulse analogs that act on these mixed \textit{frames}: for example, \path{barrier}, \path{delay}, \path{shift_phase}, \path{set_phase}, \path{shift_frequency}, \path{set_frequency}, and \path{play} are defined to sequence and modulate pulses instead of qubits.
Readout is implemented by performing a \path{play} on a readout \textit{frame} followed by a \path{capture} of that frame~\cite{ibm-qe_compiler:24}.

Note that LLVM's \gls{mlir} pass manager makes adding pulse-level support to the \gls{mqss}'s compiler straightforward by simply extending the \textit{MQSS Pass Suite} and register those passes against the appropriate dialect.
Its existing pass runner infrastructure on the other hand, allows these new passes to be seamlessly combined with gate- and pulse-based pipelines\footnote{Note that by treating pulse constructs as first-class \gls{ir} elements, LLVM frameworks like \gls{mqss} also make it possible to extend a quantum accelerator's native gate set. This means that an expert can define a new quantum gate by providing its pulse \textit{waveform} on that hardware, and the compiler will lower it into the corresponding pulse operations, seamlessly integrating the new gate into the framework.}.

Importantly, if, over time, the hardware modalities or pulse semantics require specialized behavior beyond what IBM's \textit{pulse} \gls{mlir} dialect supports, one can define a custom \textit{\gls{mqss} Pulse Dialect}.
This dialect can then leverage the same pass manager framework and \gls{mlir} infrastructure to support richer or domain-specific pulse-level semantics.

\subsection{Quantum Device Management Interface}\label{subsec:qdmi}
    \Gls{qdmi} is the hardware abstraction layer of \gls{mqss}, enabling seamless integration between software services---such as simulators, calibration tools, compilers, and telemetry-driven error mitigation---and quantum accelerators, including both physical quantum accelerators and third-party simulators~\cite{Wille:SSESSB24}.
As the \gls{qdmi} interface specification is defined in C as a header-only library, it facilitates efficient use within \gls{hpc} environments and supports services, like noise-aware simulation and automated calibration.

\begin{figure}[t!]
    \centerline{%
    \includegraphics[width=\columnwidth]{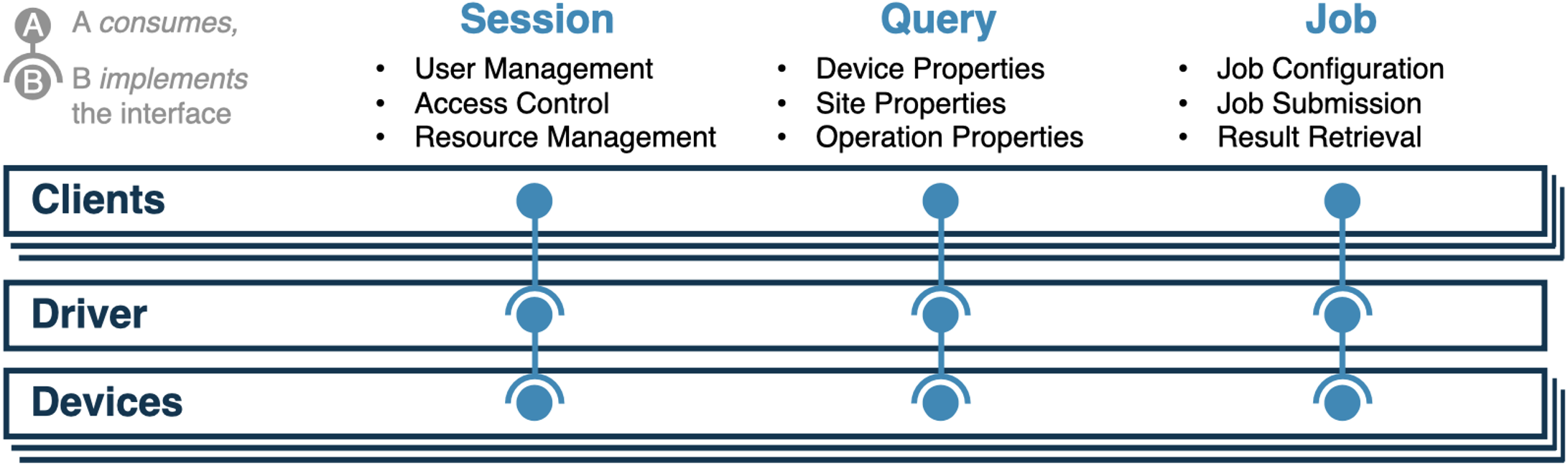}}
    \caption{An overview of the QDMI interface, its components, and how they are connected across the three QDMI entities. Adding pulse-specific device, site, and operation properties will enable clients to retrieve information---such as the level of pulse access, available channels, and more---via the existing `Query' interface. Note that submitting jobs with pulse-payload does not require modifications to the `Job' interface; it only requires adding a single enumeration value.}
    \label{fig:structure_qdmi}
\end{figure}

As seen in Fig.~\ref{fig:structure_qdmi}, \gls{qdmi} defines three primary entities:

\begin{itemize}
    \item \textbf{QDMI Clients}: The users of the QDMI library. For example, the \textit{\gls{mqss} Client}, the \textit{\gls{mqss} Compiler Passes}, or external tools.
    The clients do not have direct access to the devices but access through a \textit{QDMI Driver}.
    \item \textbf{QDMI Driver}: A bespoke solution for orchestrating these interactions, managing available \textit{\gls{qdmi} Devices} and mediating client-side requests by implementing session and job control structures.
    \item \textbf{QDMI Devices}: Quantum hardware, quantum simulators, databases, and potentially remote services provided by, for example, \glspl{isv}, and even data centers.
\end{itemize}

The \gls{qdmi} specification includes types and handlers for \textit{sites}, \textit{operations}, \textit{sessions}, and \textit{jobs}.
This novel interface uses opaque pointers and enumeration values for data structures and operations on it, ensuring that new properties or operations can be added without breaking existing interfaces.

In \gls{qdmi}, a \textit{site} references a physical or logical qubit location---e.g., a superconducting qubit, an ion-trapped qubit, or a neutral-atom trap.
\textit{Operations} encompass, for example, quantum gates, measurements, and movement primitives.
We will extend the \textit{QDMI Operations} to support pulse primitives.
In summary, to enable pulse-level support, the \gls{qdmi} specification will be extended to provide:

\begin{itemize}
    \item \textbf{Query capabilities} to see if the pulse interface is supported, and to query the types of supported pulses, their parameters and the allowed range of values. Pulse support can be provided at two levels of abstraction: \textit{site} level and \textit{port} level. 
    \item \textbf{Data structures} added to represent pulse \textit{waveforms}, pulse implementation and \textit{ports}.
    The pulse representation can be used at both \textit{site} and \textit{port} levels.
    \item \textbf{Mechanisms} to query and set default pulse implementations for specific operations, as well as to add pulse implementation for custom operations. 
    \item \textbf{A unified pulse submission interface}, to submit pulse programs in an exchange format supported by the device. 
\end{itemize}

These enhancements ensure that \gls{mqss} can query the necessary information in order to prepare and submit pulse-level payloads to quantum hardware in a standardized, backward-compatible manner.

\subsection{Exchange Format}\label{subsec:exchange_format}
    \Gls{qir} is a hardware-agnostic LLVM-compliant \gls{ir} specification that many quantum toolchains already use as a universal exchange language~\cite{stade2024supportingqirthoughtsadopting}.

Unlike textual languages such as, for example, OpenQASM, \gls{qir} can be directly compiled and linked with vendor-provided runtimes or libraries.
For instance, a \gls{qir} \textit{LLVM module} contains declared, but unimplemented quantum routines (e.g., \texttt{\_\_quantum\_\_qis\_\_h\_\_body}) that are resolved by linking in the hardware-specific definitions during execution.
As a consequence, a \gls{qir} job can become an executable intermediate object, reducing end-to-end latency and keeping the format technology- and device-agnostic.
By contrast, for example, the latest release of OpenQASM\footnote{OpenQASM 3, at the moment of writing.}---which does now include pulse constructs via \texttt{defcal} and \texttt{defcalgrammar}---is still a textual high-level format.
Likewise, IBM has deprecated its old \textit{Qiskit Pulse} schedules in favor of calibrated \textit{fractional gates} and so-called \textit{quantum dynamics} libraries.

We believe that relying on a mature, LLVM-based \gls{ir} is a future-proof solution.
Given the efficient and robust optimization routines brought by the LLVM ecosystem from the classical computing domain, state-of-the-art quantum stacks such as, for example, Quantinuum's, NVIDIA's, Rigetti's, and \gls{ornl}'s are converging on \gls{qir} or other LLVM-based \glspl{ir}~\cite{nguyen2021quantumcircuittransformationsmultilevel}.

We propose extending the \gls{qir} specification with a \textit{Pulse Profile} to natively carry pulse-level abstractions, and using that \gls{qir} with pulse support as the default exchange format for pulses in \gls{mqss} and, consequently, the \gls{qdmi} specification.
\gls{qir} already defines the notion of \textit{Profiles} to specialize this LLVM-compliant \gls{ir} for certain hardware or use cases.
Here, a so-called \textit{Pulse Profile} would augment the \textit{Base Profile} with the abstractions we introduced in Section~\ref{sec:challenges}, namely \textit{port}, \textit{frame}, and \textit{waveform}.
In practice, this could potentially mean adding new metadata and intrinsics to the \gls{qir} \textit{LLVM modules}.
For example, Listing~\ref{lst:qir_pulse} shows a human-readable \gls{qir} \textit{LLVM module} with \texttt{attributes \#0 = \{ "entry\_point" "output\_labeling\_schema" "qir\_profiles"="pulse" "re-\\quired\_num\_ports"="1" \}}.
In this snippet:

\input{listings/qir_pulse.ll}

\begin{itemize}
    \item \textbf{Pulse Profile metadata}: Attribute \texttt{qir\_profiles="pulse"} marks the \textit{LLVM module}'s entry function as using the new \textit{Pulse Profile}.
    The accompanying \texttt{output\_labeling\_schema} (and custom fields like \texttt{required\_num\_ports="1"}) signal that the program output should be formatted as a pulse job.
    These metadata align with \gls{qir}'s extensibility points---for example, the \gls{qir} specification describes \textit{Output Schemas} for calibrations or adaptive circuits---so we leverage the same mechanism to trigger a specialized pulse output format.
    \item \textbf{Opaque types matching our abstractions}: We introduce \gls{qir} opaque types \texttt{\%Port}, \texttt{\%Waveform}, and \texttt{\%Frame} (in addition to \texttt{\%Qubit} and \texttt{\%Result}) to represent the hardware elements of a pulse sequence.
    These types directly correspond to the \gls{mqss} abstractions of a control \textit{port}, \textit{frame}, and \textit{waveform} (see Sec.~\ref{sec:mqss}).
    Given that an LLVM-compliant \gls{ir} should support named opaque structs, \gls{qir} can carry these types without assuming any hardware details.
    \item \textbf{Pulse intrinsics (LLVM calls)}: Within the function (e.g. \texttt{@my\_pulse}), the code calls new \textit{pulse} intrinsics that mirror our pulse operations.
    For instance, \texttt{@\_\_quantum\_\_pulse\_\_\-waveform\_\_body(\%Waveform* \%waveform0, float* \%amps)} creates a \textit{waveform} object from amplitude samples, and \texttt{@\_\_\-quantum\_\_pulse\_\_waveform\_play\_\_body(\%Port* \%port0, \%Waveform* \%waveform0)} plays it on a \textit{port}. 
    Similarly,\\\texttt{@\_\_quantum\_\_pulse\_\_frame\_change\_\_body(\%Port* \%port0, double \%freq, double \%phase)} changes the frequency/phase on that \textit{port}, and \texttt{@\_\_quantum\_\_pulse\_\_delay\_\_\-body(\%Frame* \%frame, int)} inserts a delay.
    These are declared (but not defined) in the \textit{LLVM module}, just as the standard \gls{qir} calls are.
    At runtime, the hardware-specific \textit{\gls{qdmi} Device} layer would link these calls to the actual device \glspl{api} that implements \textit{waveform} generation and scheduling.
    \item \textbf{Integration with gate-level operations}: The example also mixes in a standard \gls{qir} \gls{qis}\footnote{That is, an instruction set compatible with the proposed \textit{Pulse Profile}.} measurement.
    After constructing and sending pulses, the code calls \texttt{\_\_quantum\_\_qis\_\_mz\_\_body(\%Qubit*, \%Result*)} to measure each qubit. This shows that pulse-level instructions can seamlessly coexist with gate-level calls in the same \gls{qir} \textit{LLVM module}.
    In \gls{mqss}, this means a single exchange file could contain the full hybrid sequence (pulses followed by measurements), preserving the program structure.
\end{itemize}

Overall, adopting a \gls{qir}-based exchange format for pulses leverages the robustness of LLVM (widely used in current stacks~\cite{Swierkowska:ESS24}) and the interoperability of a technology-neutral \gls{ir}.
\Gls{qir}'s LLVM roots mean we can apply standard compiler passes, optimizations, and linking workflows to pulse programs just as we do for classical code.
By embedding pulse semantics at the \gls{ir} level, \gls{mqss} becomes ``pulse-aware'' end-to-end while still retaining gate-level compatibility.
This then ties together the entire \gls{mqss}.

In a typical workflow, \textit{\gls{mqss} Adapters} can produce \gls{mlir}-pulse code, \gls{mqss}'s \gls{mlir}-based compiler will then \textit{lower} it to \gls{qir} with pulse support, and \gls{qdmi} will submit it to the target quantum device for the hardware runtime to execute the linked \textit{waveform} instructions.

\subsection{Consistency Across the Stack}\label{subsec:consistency}
    When combining the approach across all layers discussed above, the result is a unified design where \textit{port}, \textit{frame}, and \textit{waveform} mean the same thing at every layer, and where pulse-level access is supported in both the high-level programming interfaces, \gls{mlir}-based \gls{jit} compilation, and the lower-level \gls{qir} exchange format supported by a backend interface like, for example, \gls{mqss}'s \gls{qdmi}.

\section{Current Status}\label{sec:discussion}
    The work presented here is not a speculative vision but the result of an ongoing, coordinated effort across several fronts. 
To ground our design in hardware reality, we organized a series of day-long \glspl{tem} with providers of neutral-atom, ion-trap, and superconducting devices, with upcoming workshops planned for photonic systems.
These discussions allowed us to identify commonalities across technologies and feed those insights back into our proposal.
Concretely, \gls{qdmi} is already being extended through pull requests in its public GitHub repository~\cite{mqss-github:25}, reflecting the lessons learned from these \glspl{tem}.
In parallel, we are contributing at the ecosystem level:
\gls{lrz} is part of the \gls{qir} Alliance steering committee and successfully proposed a new workstream to extend the \gls{qir} specification with pulse-level capabilities.
This effort will build on the same hardware knowledge base while ensuring compatibility with the extensions under development for \gls{qdmi}.
On the other hand, we are analyzing how to evolve each \textit{MQSS Adapter}~\cite{mqss-adapters-github} to accept pulse-level job definitions in a manner that remains compatible with existing pulse programming interfaces---thereby avoiding integration issues should \gls{mqss} add support for those platforms in the future as well.
Once the \textit{MQSS Adapters} are updated, the \textit{MQSS Client} will be modified to forward this payload to the rest of the stack.
Finally, because the \gls{mqss} compiler is LLVM/\gls{mlir}-based and dialect-agnostic, we expect only minimal core changes; nevertheless, we will empirically evaluate whether a dedicated \gls{mqss} pulse \gls{mlir} dialect is needed as implementation experience accumulates.

\section{Related Work}\label{sec:related_work}
    Pulse-level control appears in several mainstream frameworks.
Qiskit-Pulse (already deprecated) and Amazon Braket Pulse, for example, expose the same core abstractions we identified, that is, \textit{ports}, \textit{frames}, and \textit{waveforms}, via Python \glspl{api}, providing convenient primitives for \textit{waveform} construction, \textit{frame} management, and scheduling. 
These interfaces are useful for experimentation, but their Pythonic nature and limit suitability for low-latency, tightly integrated \gls{hpc} workflows.

On the \gls{ir} and serialization side, OpenQASM~3 (with its calibration/\texttt{cal} blocks) and legacy formats such as \texttt{Qobj} demonstrate how pulse schedules can be described and exchanged, yet they remain tied to their originating ecosystems and are not designed as compiled, linkable \glspl{ir}.
Several groups (including at Quantinuum and ORNL~\cite{nguyen2021quantumcircuittransformationsmultilevel}) are moving their compiler toolchains toward LLVM/\gls{mlir}-based designs to gain the benefits of compiled passes and tighter hardware integration; however, these projects typically address only parts of the stack (\gls{ir} or runtime) rather than offering an end-to-end, \gls{hpc}-centric solution.

For resource and device integration, the \gls{qrmi} introduced in~\cite{qrmi:25} and similar proposals address \gls{hpcqc} integration, solving lifecycle and access-control problems, but not the compiler-level and pulse-format challenges required for native pulse programs.
\gls{mqss} (with its \gls{qpi}, LLVM/\gls{mlir} compiler, \gls{qdmi}, and the proposed \gls{qir} with pulse support as exchange format) differs by targeting all four layers at once: a compiled, low-latency programming \gls{api}, dialect-aware compilation, and a C/C++ backend query/management interface along with a \gls{qir}-based pulse exchange mechanism that align the same \textit{port}, \textit{frame}, and \textit{waveform} abstractions across the entire pipeline.

That is, prior systems address important subproblems, but none yet combine compiler-aware \gls{ir}, low-overhead programming \gls{api}, backend interfaces, and a portable pulse exchange format in a single \gls{hpc}-oriented stack, the gap our proposals is designed to fill.

\section{Conclusions}\label{sec:conclusions}
    We presented a detailed discussion of the obstacles to integrating pulse-level quantum control into an \gls{hpc} stack and proposed solutions within \gls{mqv}'s \gls{mqss}. 
Our analysis identified three required pulse abstractions, namely, \textit{ports}, \textit{frames}, and \textit{waveforms}, to be supported at programming interface, \gls{ir}, backend interface, and exchange format levels. 
We also introduced the appropriate extensions to \gls{mqss}.
Concretely, we 1) introduced an extension to its C-based programming \gls{api} with pulse constructs to avoid Python runtime overhead, 2) illustrated the adoption of an \gls{mlir} pulse dialect to represent pulse commands alongside gate-level instructions, 3) leveraged its C/C++ quantum backend interface specification to query hardware pulse constraints during \gls{jit} compilation, and 4) defined an extension to add pulse-level support to the \gls{qir} specification and adopt it as an exchange format to leverage dynamic pulse constructs linking to accelerator implementations.
These adaptations  allow \gls{mqss} and other similar heterogeneous \gls{hpcqc} software stacks to natively represent and compile pulse sequences, while remaining compatible with existing \gls{hpc} scheduling and execution models.
Our work establishes an end-to-end path for pulse-aware hybrid quantum-classical workloads: by embedding the low-level pulse semantics into each layer of the software stack, we enable advanced control techniques (calibrations, custom \textit{waveforms}, etc.) within \gls{hpc} environments.
This pulse-enabled \gls{hpcqc} stack opens the door for new kinds of quantum-accelerated algorithms and more effective exploitation of near-term hardware.

\begin{acks}\label{sec:acknowledgment}
    This work is supported by the German Federal Ministry of Research, Technology and Space (BMFTR) with the grants 13N15689 (DAQC), 13N16063 (Q-Exa), 13N16078 (MUNIQC-Atoms), 13N16187 (MUNIQC-SC), 13N16690 (Euro-Q-Exa), 13N16894 (MAQCS), European fundings 101136607 (CLARA), 101114305 (Millenion), 10111394\-6 (OpenSuperQPlus), 101194491 (QEX), and the Bavarian State Ministry of Science and the Arts (StMWK) through funding, as part of \gls{mqv}, Q-DESSI.
\end{acks}

\bibliographystyle{ACM-Reference-Format}
\setcitestyle{sort}
\bibliography{bibliography}

\end{document}